%% file: main.tex
\documentclass[sigconf]{acmart}

\usepackage{algorithm}
\usepackage{algpseudocode}
\newcommand{\algname}[1] {{\fontfamily{cmtt}\selectfont {#1}}}
\usepackage{dsfont}

\usepackage{caption}
\usepackage{subcaption}
\usepackage{multirow}
\usepackage{graphicx,wrapfig,lipsum}

\usepackage{float}
\usepackage[absolute]{textpos}

\AtBeginDocument{%
  }

\setcopyright{none}
\renewcommand\footnotetextcopyrightpermission[1]{}
\begin{document}

\begin{textblock}{10}(3,0.4)
 \noindent\normalsize  \begin{center}5th FAccTRec Workshop on Responsible Recommendation \\ in conjunction with ACM RecSys 2022\end{center}
 \end{textblock}

\title{Exposure-Aware Recommendation using Contextual Bandits}

\author{Masoud Mansoury}
\email{m.mansoury@uva.nl}
\affiliation{%
  \institution{University of Amsterdam}
  \institution{Discovery Lab, Elsevier}
  \city{Amsterdam}
  \country{Netherlands}
}
\author{Bamshad Mobasher}
\email{mobasher@cs.depaul.edu}
\affiliation{%
  \institution{DePaul University}
  \city{Chicago}
  \country{United States}
}
\author{Herke van Hoof}
\email{h.c.vanhoof@uva.nl}
\affiliation{%
  \institution{University of Amsterdam}
  \city{Amsterdam}
  \country{Netherlands}
}

\renewcommand{\shortauthors}{Mansoury et al.}

\begin{abstract}
  Exposure bias is a well-known issue in recommender systems where items and suppliers are not equally represented in the recommendation results. This is especially problematic when bias is amplified over time as a few items (e.g., popular ones) are repeatedly over-represented in recommendation lists and  users' interactions with those items will amplify bias towards those items over time resulting in a \textit{feedback loop}. This issue has been extensively studied in the literature on model-based or neighborhood-based recommendation algorithms, but less work has been done on online recommendation models, such as those based on top-$K$ contextual bandits, where recommendation models are dynamically updated with ongoing user feedback. In this paper, we study exposure bias in a class of well-known contextual bandit algorithms known as \textit{Linear Cascading Bandits}. We analyze these algorithms on their ability to handle exposure bias and provide a fair representation for items in the recommendation results. Our analysis reveals that these algorithms tend to amplify exposure disparity among items over time. In particular, we observe that these algorithms do not properly adapt to the feedback provided by the users and frequently recommend certain items even when those items are not selected by users. To mitigate this bias, we propose an \textbf{E}xposure-\textbf{A}ware (EA) reward model that updates the model parameters based on two factors: 1) user feedback (i.e., clicked or not), and 2) position of the item in the recommendation list. This way, the proposed model controls the utility assigned to items based on their exposure in the recommendation list. Extensive experiments on two real-world datasets using three contextual bandit algorithms show that the proposed reward model reduces exposure bias amplification in long run while maintaining the recommendation accuracy.
\end{abstract}

\ccsdesc[500]{Information systems~Users and interactive retrieval}
\ccsdesc[500]{Information systems~Recommender systems}

\keywords{recommender systems, contextual bandits, exposure fairness}

\maketitle

\section{Introduction}

Recommender systems utilize users' interaction data on different items to generate personalized recommendations \cite{resnick1997recommender,jannach2010recommender}. In online interactive recommender systems, part of the users' interaction data come from the users' feedback on items shown in the recommendation list generated by the recommendation model. Users may or may not select recommended items: clicked or selected items are considered as the positive samples and unclicked items are considered as the negative samples. The recommendation model uses these positive/negative signals to learn users' preferences. This process is the main paradigm used in contextual bandit algorithms where the interactions between users and recommendation system over time is used to build the recommendation model \cite{joseph2016fairness,wu2016contextual}.


Various contextual bandit algorithms have been proposed as the basis for online recommendation \cite{lattimore2018toprank,li2019online,zoghi2017online,li2016contextual,yue2011linear}. These algorithms solely focus on learning users' preferences to increase the click-through rate of the recommendations. However, this user-centric view for building recommendation model neglects the item-side utilities. Exposure to items in the recommendation results is an important item-side utility that can have direct influence on economic gain for items and their suppliers. Bias in exposure could lead to unfair treatment of some suppliers potentially resulting in a disincentive to participate in the market. It may also inhibit the ability of the system to provide useful, but less popular recommendations to consumers. Hence, the question is how contextual bandit algorithms distribute exposure among items in the system? Although these algorithms perform exploration in the items space to collect users' feedback on different items, our study in this paper shows that this exploration does not necessarily lead to a fair exposure for items in the long run.

   
\textit{Exposure bias} in recommender systems is a well-known problem that refers to the fact that items are not uniformly represented in the recommendation results: few items are frequently shown in the recommendation lists, while majority of other items rarely appear in the recommendation results \cite{abdollahpouri2020multi}. This phenomenon may result in an unfair representation of items from different suppliers in recommendations. Various notions of exposure bias (or its counterpart, exposure fairness) are introduced in the literature: 1) aggregate diversity which is defined as the fraction of items that appeared at least once or $\alpha$ times in the recommendation lists \cite{mansoury2021understanding,antikacioglu2017post,adomavicius2011maximizing,mansoury2021graph,mansoury2020fairmatch}, 2) equality of exposure which is defined as how equally the total exposure is distributed among all items \cite{mansoury2021understanding,marras2021equality,polyzou2021faireo,mansoury2021graph}, and 3) equity of exposure which is defined as how equally the total exposure is distributed among all items in the system with respect to their qualification/relevance, meaning that high-quality items (based on users' interaction) are expected to receive higher exposure than the low-quality ones \cite{wang2021fairness,singh2019policy,singh2018fairness}. In this paper, we consider the second definition and introduce several evaluation metrics for evaluating the performance of the recommendation system in long run.   

Most existing research to study exposure bias are conducted in static settings where a single round of recommendation results is analyzed \cite{patro2020fairrec,mansoury2021understanding,suhr2019two}. While these studies unveiled important aspects of exposure bias and proposed solutions for tackling it, the long-term impact of this bias is yet a significant research gap which we seek to investigate in this paper. Filling this gap requires studying the task of recommendation problem in dynamic and interactive settings where users are engaged in ongoing interaction with the system and preference models are dynamically updated over time. Our choice of contextual bandit algorithms meets this requirement as these algorithms operate in a dynamic online recommendation environment.    

Due to the feedback loop phenomena in which  the users and the recommender system are in a process of mutual dynamic evolution
, exposure bias, if not mitigated, can be even intensified over time \cite{mansoury2020feedback,sinha2016deconvolving}. Highly exposed items in the recommendation lists have higher chance to be viewed/examined/clicked by the users, while insufficiently exposed items will not receive proportionate attention from the users. As a result, those over-exposed items would have a higher chance to be shown in the recommendation lists in the future which amplifies existing bias. Amplification of exposure for few items would be at the expense of the under-exposure for a majority of other items (the ones that might be interesting for some users) and consequently may make those items completely out of the market. 

In this paper, we study exposure bias in contextual bandit algorithms and the degree to which these algorithms are vulnerable to exposure bias. Our analysis shows that these algorithms are unable to fairly assign exposure to different items and in long run, a huge disparity is observed in the exposure assigned to different items. We understood that the main reason for this unfair behavior is that the contextual bandit algorithms do not properly adapt to the feedback provided by the users. We observe that few items are frequently shown in the recommendation lists even though in majority of cases the users do not click on them. That is, although those few items make large false positive, the recommendation model repeatedly shows them in the recommendation lists. 

To overcome this deficiency, we propose an \textit{Exposure-Aware reward model} and integrate it into the existing contextual bandit algorithms. Our algorithm updates the model parameters based on two factors: 1) the user feedback on recommended items, whether item is clicked or not, and 2) the position of the item in the list. In fact, the proposed model rewards or penalizes the clicked or unclicked items, respectively, based on their position in the recommendation list. When an item recommended at the bottom of the list is clicked by the user, then it would be significantly rewarded for that user. On the other hand, if an item recommended on top of the list is not clicked by the user, then it would be significantly penalized for that user. This control over the degree of reward or penalization for the items based on their exposure in the recommendation lists helps to better adapt to the users' preferences and reduces exposure bias on items. 

Extensive experiments using three contextual bandit algorithms on two publicly-available datasets show that the proposed reward model improves the exposure fairness for items, while maintaining the accuracy of the recommendations. Unlike the existing studies in the literature that have shown a trade-off between improving the recommendation accuracy for users and fairness of exposure for items \cite{singh2019policy,jeunen2021top}, in this work, we show that taking into account the item exposure in the model and properly adapting the model based on the users' feedback not only improves exposure fairness for items, but also in some cases leads to increase in the number of clicks on the recommended items.

The following are our main contributions:

\begin{itemize}
    \item Through empirical study and simulating recommendation system in dynamic online setting, we investigate the well-known exposure bias problem in contextual bandit algorithms. We find that those algorithms are negatively affected by exposure bias and even amplify bias over time. 
    \item We propose an exposure-aware reward model that adapts the model based on users' feedback on recommended items, but also take into account the degree of exposure for each of those recommended items. 
    \item Finally, through extensive experiments on two datasets, we show that integrating the proposed exposure-aware reward model into the existing contextual bandit algorithms improves the exposure fairness for items, while maintaining recommendation accuracy.  
\end{itemize}

\section{Background}

Let us denote $\mathcal{U}=\{u_1,u_2,...,u_n\}$ as a set of $n$ users and $\mathcal{I}=\{i_1,i_2,...,i_m\}$ as a set of $m$ items in the system. We formulate the recommendation problem as the ranking of $m$ candidate items into $K$ positions where $K\leq m$. We denote the recommendation list delivered to each user $u$ as $R_u$.  

\subsection{Cascade feedback model}\label{cm}

Cascade model \cite{craswell2008experimental} is used for modeling user’s click behavior. In this model, the user examines each recommended item one-by-one from the first position to the last, clicks on the first attractive item, and does not examine the rest of the items. This way, the items above the clicked item are considered unattractive, the clicked item is considered as attractive, and the rest of the items are considered as unobserved (neither attractive nor unattractive). Cascade model is effective in addressing well-known \textit{position bias} \cite{collins2018position,hofmann2014effects} where lower ranked items in the recommendation list are less likely to be clicked than the higher ranked items. By not considering the lower ranked (unobserved) items as unattractive, the model will not consider them as the negative or unattractive items and may give them chance to appear in the higher rank in the future.

\subsection{Linear cascading bandit algorithms}\label{cb}

Contextual bandit algorithms are online learning-to-rank algorithms that learn users' preferences by iteratively interacting with the users. These algorithms utilize the information about the items' contents (e.g. latent representations or explicit features) in the existing bandit algorithms (e.g. Thompson Sampling \cite{agrawal2012analysis}, Upper Confidence Bound \cite{auer2002finite}) to learn users' preferences. Various contextual bandit algorithms are developed in the literature \cite{lattimore2018toprank,li2019online,zoghi2017online,li2016contextual,yue2011linear} by incorporating contextual information into existing bandit algorithms. One class of these algorithms is Linear Cascading Bandits \cite{zong2016cascading,hiranandani2020cascading,li2020cascading} that combine a bandit algorithm with contextual data and cascade feedback model, described above, for collecting users' feedback. Different bandit algorithms including Thompson Sampling and Upper Confidence Bound (UCB) are used under this setting to develop a contextual bandit algorithm \cite{zong2016cascading}. In this paper, we focus on UCB-style cascading bandit algorithms \cite{zong2016cascading,hiranandani2020cascading,li2020cascading} and propose ideas to further improve the performance of these algorithms. 

        \begin{algorithm}[t!]
        \caption{Top-$K$ Contextual Bandit Recommendation Algorithm}\label{alg:cap}
        \begin{algorithmic}[1]
        \Require Number of rounds $T$, size of recommendation list $K$
        \State Initialize model parameters $\mathcal{M}$
        \For{$t=1,...,T$}
            \State Observe context $\theta_t$
            \State Generate recommendation list of size $K$ according to $\mathcal{M}$
            \State Observe user feedback on recommended items
            \State Update model parameters $\mathcal{M}$
        \EndFor
        \end{algorithmic}
        \end{algorithm}

In linear cascading bandits, the learning agent interacts with the users by delivering the recommendations to them and receiving feedback. The users' feedback is modeled using cascade model and based on the feedback from the users, the learning agent updates its model. In the following, we elaborate each steps of learning process in cascading bandits as shown in Algorithm \ref{alg:cap}.

\subsubsection{Observing context.} To generate the recommendation lists, the learning agent needs to compute the probability for each item that a target user may like. This probability is called \textit{attraction probability} and is computed as $W=X\theta^{*T}$ where $X\in \mathbb{R}^{m \times d}$ is the known $d$-dimensional matrix of items features, $\theta^* \in \mathbb{R}^{1 \times d}$ is the $d$-dimensional vector of embedding (i.e. representation of preferences) for the target user $u \in \mathcal{U}$, and $W \in \mathbb{R}^{m \times 1}$ is the attraction probability of each item for $u$. Since $\theta^*$ is unknown and needs to be learned by interacting with the user, the representation of user preferences needs to be estimated by solving a ridge regression problem on items embedding and the attraction probabilities over the past observations in $t-1$ rounds. Considering the item embedding as independent variables and the attraction probabilities as dependent variable, the representation of a target user preferences can be estimated as:
\begin{equation}
    \hat{\theta}_t = (X^TX+ \lambda I)^{-1}X^TW_{t-1}
\end{equation}

We denote $M_t=X^TX+ \lambda I$ as the co-variance matrix of item features and $B_t=X^TW_{t}$ as the importance weight of item features for target user at round $t$. Both $M \in \mathbb{R}^{d \times d}$ and $B \in \mathbb{R}^{d \times 1}$ are the model parameters and their values are updated based on user's feedback on recommended items.  

\subsubsection{Recommendation generation.} In this step, a score for each item is computed and then top-$K$ items with the highest scores are returned as the recommendation list for the target user. According to UCB, this score is computed by combining the estimated attraction probability of each item $i$ and the upper confidence bound for this estimation:
\begin{equation}\label{u_t}
    U_t(i) = x_i\hat{\theta}_{t-1}^T + c\sqrt{x_iM^{-1}_{t-1}x_i^T}
\end{equation}

\noindent where the term $\sqrt{x_iM^{-1}_{t-1}x_i^T}$ is the upper confidence bound for the estimated attraction probability of item $i$ that covers the optimal attraction probability and is computed by the norm of $x_i$ weighted by $M^{-1}$ (i.e. $||x_i||_{M^{-1}}$). This upper confidence bound encourages the exploration in the items space and the hyperparameter $c$ controls the degree of exploration. Given a score computed for each item, $K$ items with the highest $U_t(i)$ are returned to form the ordered recommendation list $R_u$ for user $u$.

\subsubsection{Collecting user click.} Given recommendation lists $R_u$, the agent receives feedback $C_{t,u}$ from target user $u$ which is the index of the clicked item in $R_u$. Thus, if $C_{t,u} \leq K$, the user clicked on item at position $C_{t,u}$, while if $C_{t,u}=\infty$, then user did not click on any item. According to cascade click model, the observed weights of all recommended items would be updated as $w_{i_k}=\mathds{1}(C_{t,u}=k)$ for all $k=1,...,min\{C_{t,u},K\}$.

\subsubsection{Updating model parameters.} After receiving user feedback, the agent updates the model parameters: $M$ and $B$. For each examined item $e$ (i.e. $e \in \{i_k:k \leq min\{C_{t,u},K\}\}$, $M$ and $B$ are updated as follows:
\begin{equation}\label{update_M}
    M_t \leftarrow M_t + x_e^Tx_e
\end{equation}
\begin{equation}\label{update_B}
    B_t \leftarrow B_t + x_ew_e 
\end{equation}

In practice, for improving the efficiency of the algorithm, $M^{-1}$ is updated instead of $M$ as computing the inverse of $M$ at each round is computationally expensive \cite{zong2016cascading}.  



\section{Formalizing Item Exposure in Dynamic Recommendation Setting}\label{exposure}

In most of the existing works, exposure of an item is measured as the number of times that item appeared in the recommendation lists of different users (e.g., aggregate diversity) \cite{mansoury2021understanding,antikacioglu2017post,adomavicius2011maximizing}. This definition does not consider the position of the items in the list. Items appeared on top of the list have higher chance to be examined by the user than the items at the bottom of the list. It is possible an item gets recommended, but receives no exposure because it was probably shown at the bottom of the list and the user has not examined it. For this reason, items appearing on top of the list have much higher exposure than items appeared at the bottom of the list. Therefore, each position in the recommendation list has certain exposure value. 

In other works \cite{singh2018fairness,singh2019policy,wang2021fairness}, even though the position of the items in the recommendation lists is taken into account for measuring the exposure value for an item, they are not suitable for measuring the exposure value in dynamic recommendation settings. This means that those definitions are solely designed for static recommendation settings where exposure of the items can only be computed in a single step of recommendation results. However, recommendation system is a dynamic environment where its performance at time $t$ can be affected by its output at time $t-1$, what is known as \textit{feedback loop} phenomenon \cite{zhu2021popularity,mansoury2020feedback,sinha2016deconvolving}. Also, due to the fact that recommendation slots are limited at each time step (i.e., only $K$ items can be recommended to each user), it is possible that an item does not get enough exposure at time $t$, but this under-exposure might be compensated at time $t+1$. Therefore, computing the exposure in a static recommendation setting may not accurately show the amount of exposure given to each item. A proper definition for item exposure is the one that: 1) considers position of the item in the recommendation list, and 2) computes the exposure value in a dynamic recommendation setting.

In this section, we formalize the exposure of an item in dynamic recommendation setting by considering the number of times the item is recommended and the position of that item in the recommendation list. We assume that the system is operating for $T$ rounds and at each round, a recommendation list of size $K$ is generated for each user. Then, exposure of an item $i$ based on the number of times $i$ appeared in the recommendation lists can be computed as:
\begin{equation}\label{e}
    E_i = \sum_{t=1}^{T}{\sum_{u \in \mathcal{U}}\mathds{1}(i \in R_u)}
\end{equation}

To incorporate the position weight for each item in the list, following the idea in \cite{singh2018fairness}, we consider a standard exposure drop-off (i.e., position bias) as commonly used in ranking metrics (e.g., nDCG) as follows:
\begin{equation}\label{pe}
    PE_i = \sum_{t=1}^{T}{\sum_{u \in \mathcal{U}}\sum_{k=1}^{K}\mathds{1}(i = R^k_u).\frac{1}{\log_2(1+k)}}
\end{equation}
\noindent where $PE_i$ is the \textit{position-based exposure} of item $i$, $R^k_u$ is $k$-th recommended item in $R_u$, and the term $1/\log_2(1+k)$ assigns weight to each position $k$ in the recommendation list, higher weight to items on top of the list and lower weight to the items at the bottom of the list. We denote normalized $PE_i$ as $NPE_i$ and define it as:
\begin{equation}\label{npe}
    NPE_i=\frac{PE_i}{\sum_{i \in \mathcal{I}}{PE_i}}
\end{equation}

The exposure definition in equation \ref{pe} is based on the appearance of the item in the recommendation lists and does not take into account whether or not the item is examined by the user. Another way for defining exposure of an item is based on examination of that item by the user: item has been recommended to the user and user examined it. That is, the exposure for an item would be counted when the item is examined by the user. Thus, the definition of exposure for an item $i$ can be modified as follows:
\begin{equation}\label{pee}
    PEE_{i} = \sum_{t=1}^{T}{\sum_{u \in \mathcal{U}}\sum_{k=1}^{\min(C_{t,u},K)}\mathds{1}(i = R^k_u).\frac{1}{\log_2(1+k)}}
\end{equation}
\noindent where $PEE_i$ is the \textit{position-based examined exposure} of item $i$ and $C_{t,u}$ is the index of the item clicked by $u$ in $R_u$ (e.g., if $u$ clicked on an item at position 5, then $C_{t,u}=5$). We denote normalized $PEE_i$ as $NPEE_i$ and define it as: 
\begin{equation}\label{npee}
    NPEE_i=\frac{PEE_i}{\sum_{i \in \mathcal{I}}{PEE_i}}   
\end{equation}

The exposure definitions in equations \ref{pe} and \ref{pee} cumulatively measure the amount of exposure given to each item in $T$ rounds. Both definitions take into account the number of times item appeared in the recommendation lists and the position of the item in the lists. Also, they are suitable for measuring the exposure of the items in a dynamic recommendation setting.

\section{Empirical study of exposure bias in online recommendation}

In this section, we perform sets of experiments to study exposure bias in dynamic recommendation setting using three contextual bandit algorithms on two datasets. We aim at understanding to what extend these algorithms are vulnerable to exposure bias and the potential reasons for bias amplification in the long run. 
For the experiments, we follow the experimental setting and data
preprocessing used in \cite{hiranandani2020cascading,li2020cascading}. 

\subsection{Experimental setup}\label{exp_setup}

Evaluation of interactive recommendation algorithms is usually done by off-policy evaluation approaches which does not require any online experiments  \cite{li2010contextual,zhan2021off,wang2017optimal}. However, due to the fact that in our problem the action space is too large (i.e. exponential in $K$) for commonly used off-policy evaluation methods, we utilize a simulated interaction environment for our evaluation where the simulator is built based on offline datasets. This is the evaluation setup used in most of the research works \cite{li2019online,zoghi2017online,li2016contextual,yue2011linear,zong2016cascading,hiranandani2020cascading,li2020cascading}.

\noindent\textbf{Datasets.}
We perform our experiments on two publicly available datasets: Amazon Book 
\cite{ni2019justifying} and MovieLens 
\cite{harper2015movielens}. On Amazon Book dataset, 15M users provided 49M ratings on 3M books. 
On MovieLens dataset, ~6K users provided ~1M ratings on ~4K items. 
On both datasets, the ratings are on a 5-star rating scale, each item (book on Amazon Book and movie on MovieLens) is assigned at least one genre, and genres are considered as topics.

\noindent\textbf{Data preprocessing.} We follow the data preprocessing approach in \cite{zong2016cascading,hiranandani2020cascading,li2020cascading,li2019online}. First, on both datasets, we map the ratings onto a binary scale: rating 4 and 5 are converted to 1 and other ratings to 0. Then, on Amazon Book dataset, to reduce the size and the sparsity of the dataset, we first create a core-100 sample and, then extract 1K most active users and 2K most rated items from the user-item interaction data. In this sample, there are 307 unique genres assigned to different books. On MovieLens dataset, we create a sample of this dataset by extracting 1K most active users from the user-item interaction data. In this sample, there are 18 unique genres assigned to the movies.  

\noindent\textbf{Contextual bandit algorithms.} The algorithms studied in this paper generally follow the process described in section \ref{cb} and only differ in how they create item feature matrix ($X$) and generate recommendations at step 2 of Algorithm 1. 

\begin{figure*}[t!]
    \centering
    \begin{subfigure}[b]{0.8\textwidth}
        \includegraphics[width=\textwidth]{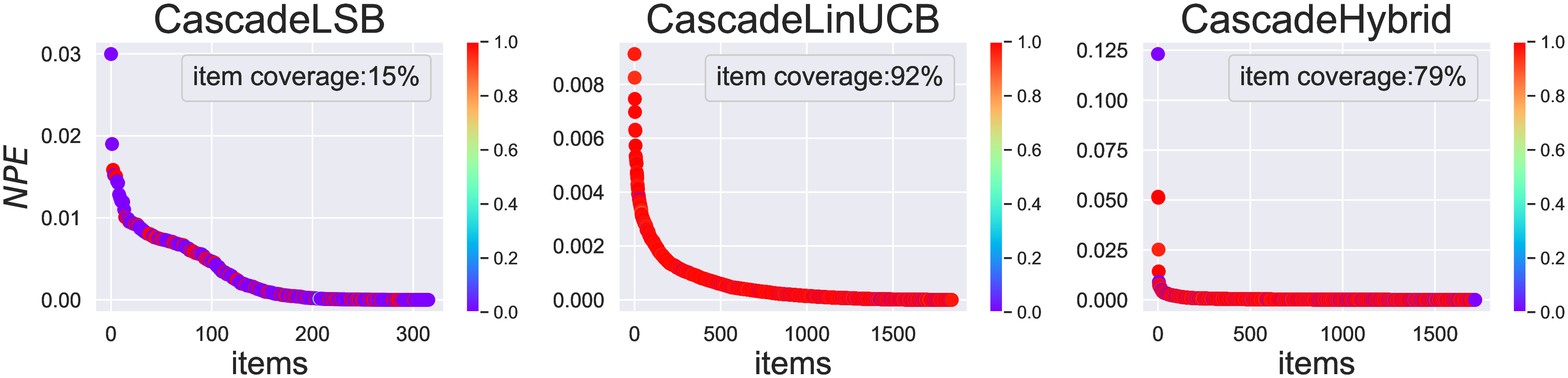}
        \caption{Amazon Book}\label{}
    \end{subfigure}
    \begin{subfigure}[b]{0.8\textwidth}
        \includegraphics[width=\textwidth]{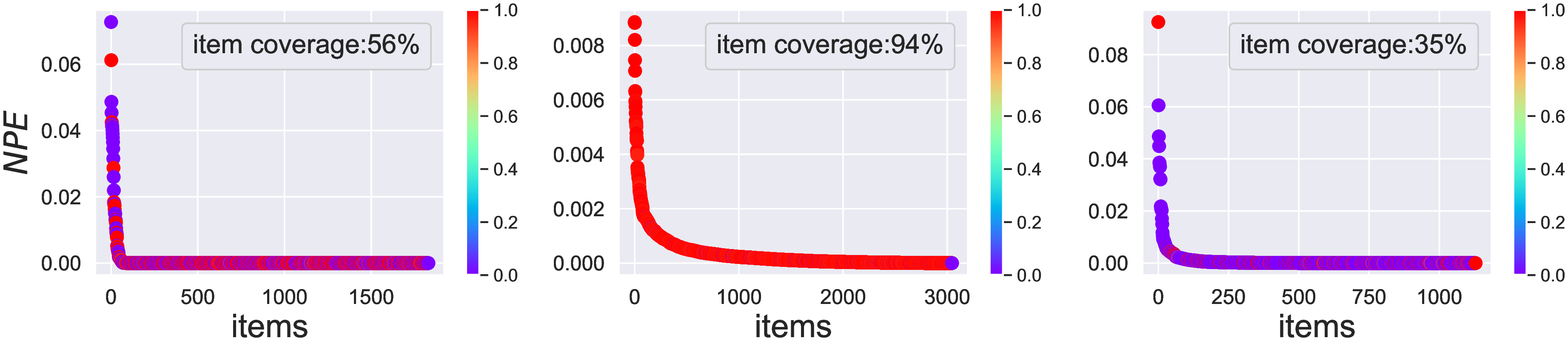}
        \caption{MovieLens}\label{}
    \end{subfigure}
\caption{Exposure distribution of items in recommendation lists in $T$ rounds. Exposure of items are computed using equation \ref{npe} and items are sorted by their exposure value in descending order. The colorbar shows the false positive rate (i.e., recommended but not clicked) of each item. Item coverage shows the fraction of items shown at least once in the recommendation lists of $T$ rounds.}\label{fig_npe_dist}
\end{figure*}

\begin{itemize}
    \item \algname{CascadeLSB} \cite{hiranandani2020cascading}: This algorithm optimizes to generate diversified recommendations. Given $X$ as the item-topic matrix, it generates recommendations to contain items from diverse topics. For this purpose, the notion of topic coverage of an item is used to describe the probability that the item covers each topic. Similarly, the topic coverage of a list is defined as the probability of the items in the list covering each topic. Assuming one or more topics are associated with each item and topic coverage for each item, the recommendation list is generated by iteratively adding items that increase the gain in topic coverage of the recommendation list. Thus, the feature vector of each item, $x_i$, is defined as the gain in topic coverage by adding that item to the recommendation list created so far. This way, the feature vector for each item is dependent to the items already added to the list. If the target item is diverse in terms of topics with the items previously added to the recommendation list, the target item will have higher chance to be added to the list.  
    \item \algname{CascadeLinUCB} \cite{zong2016cascading}: This algorithm optimizes to generate the recommendations to be relevant to the users preferences. In this algorithm, the matrix of item features, $X$, is derived by performing singular-value decomposition on user-item interaction data and the recommendations are generated by selecting the most relevant items to the users' preferences (i.e., top-$K$ items with the highest $U_t$ in equation \ref{u_t}).
    \item \algname{CascadeHybrid} \cite{li2020cascading}: This algorithm combines \algname{CascadeLSB} and \algname{CascadeLinUCB} for generating recommendations that are both relevant to the users’ preferences and diverse in terms of topics.
\end{itemize}

\begin{figure*}[t!]
    \centering
    \begin{subfigure}[b]{0.8\textwidth}
        \includegraphics[width=\textwidth]{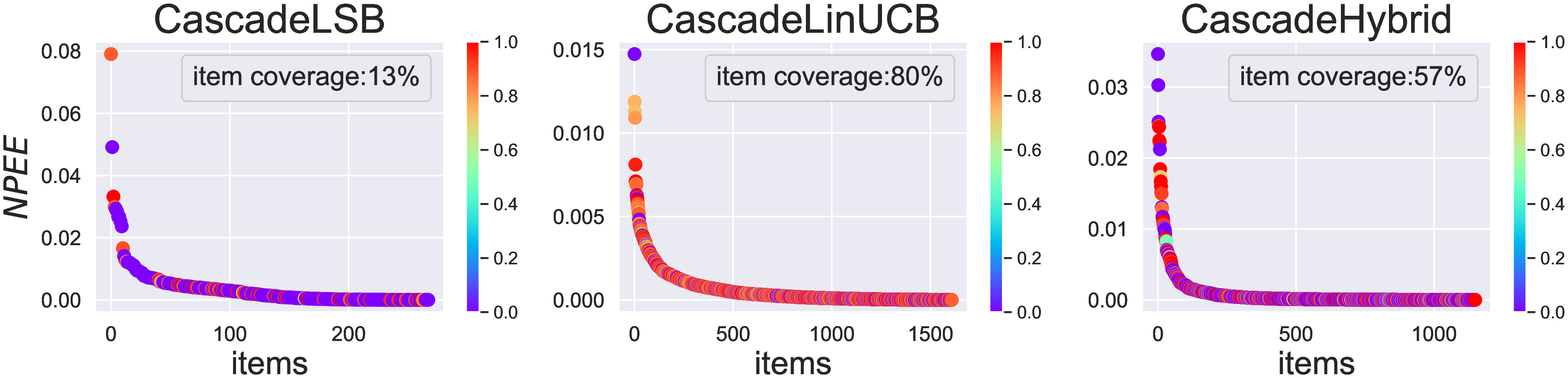}
        \caption{Amazon Book}\label{}
    \end{subfigure}
    \begin{subfigure}[b]{0.8\textwidth}
        \includegraphics[width=\textwidth]{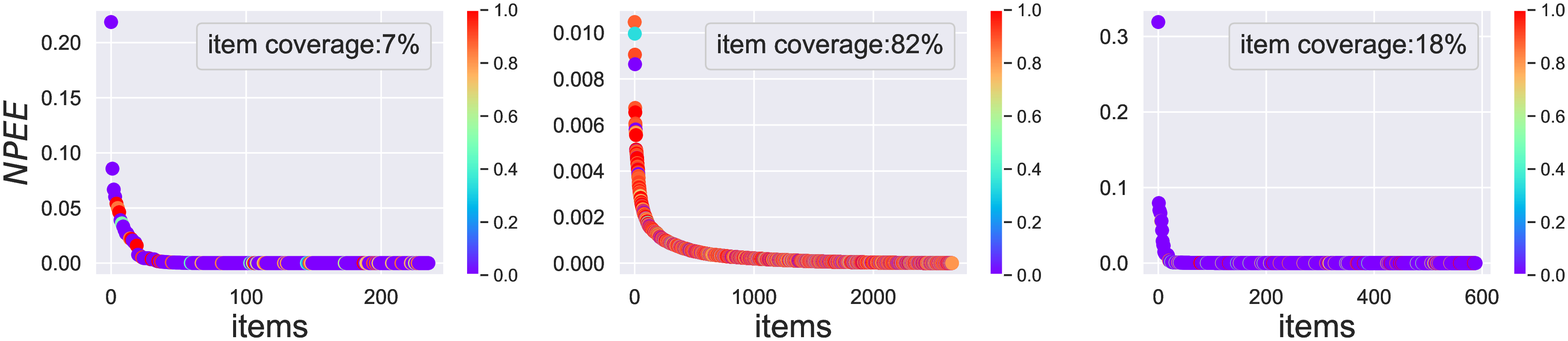}
        \caption{MovieLens}\label{}
    \end{subfigure}
\caption{Exposure distribution of examined items in recommendation lists in $T$ rounds. Exposure of items are computed using equation \ref{npee} and items are sorted by their exposure value in descending order. The colorbar shows the false positive rate (i.e., recommended and examined but not clicked) of each item. Item coverage shows the fraction of items examined at least once by the users in $T$ rounds.}\label{fig_npee_dist}
\end{figure*}

\noindent\textbf{Simulation.} We follow the simulation process in \cite{hiranandani2020cascading,li2020cascading}. For simulating the interaction between the learning agent and the users, we randomly divide users profile into 50\% as training set and 50\% as test set\footnote{This splitting ratio was also used in previous research works \cite{zong2016cascading,hiranandani2020cascading,li2020cascading} and the assumption is that even with small training data (e.g., 50\% or less) as a prior knowledge, the bandit algorithm would be able to interactively learn the users' preferences.}. The training set is used for computing the attraction probability of each item and generating the recommendation list to each user. Test set is used for modeling user feedback on recommendation list and generating the optimal recommendation list for evaluating the performance of the model. 

The assumption in linear cascading bandits is that the matrix $X$ (item-feature matrix in \algname{CascadeLinUCB}, item-topic matrix in \algname{CascadeLSB}, and a combination of item-feature and item-topic matrices in \algname{CascadeHybrid}) is known and user's feature vector is unknown which is interactively learned by the learning agent. Thus, $X^{train}$ is derived from training data and is used for generating the recommendation list, $R_u$, for the target user $u$ at each iteration. Test data is used to derive $X^{test}$ and $\theta^*$ as preference vector for $u$ to evaluate $R_u$ and to determine which item in $R_u$ would be clicked by $u$. That is, for each item $i \in R_u$, the probability of clicking $i$ is computed by $x_i^{test}.\theta^{*T}$.\footnote{Test data is also used to generate the optimal recommendation list, $R^*_u$, for measuring how close the outputs from bandit algorithm is from the optimal ranker.}


Each linear cascading bandit algorithm described in section \ref{exp_setup} uses a specific optimal ranker for generating the optimal recommendation list. In this paper, for the purpose of consistency, we use the optimal ranker of \algname{CascadeHybrid} which is the combination of \algname{CascadeLSB} and \algname{CascadeLinUCB} to generate the optimal recommendation lists. 

We perform the experiments for $T=50K$ rounds and at each round, recommendation list of size $K=20$ is generated for one randomly selected user. This is what usually happens in real-world scenarios where at each time step, a user visits the platform and the system shows the recommendation list to her.

\subsection{Experimental results}

The first question to investigate is: how are items represented in the recommendation lists generated in $T$ rounds of running recommendation system? Figure \ref{fig_npe_dist} shows the distribution of item exposure in $T$ rounds. Horizontal axis is the items sorted by their exposure value in descending order\footnote{Item rank is shown as the item ID} and vertical axis is the item exposure computed by equation \ref{npe}. Only items with $NPE > 0$ are shown in the plots. Item coverage in each plot shows the fraction of items that appeared at least once in the recommendation lists of $T$ rounds by the corresponding contextual bandit algorithm ($NPE_i=0$ means that $i$ never appeared in the recommendation list). As shown, in all algorithms on both datasets, certain items are never shown in the recommendation lists. \algname{CascadeLSB} has the lowest item coverage with 15\% on Amazon Book and \algname{CasacdeHybrid} with 35\% on MovieLens. Also, it is evident that the exposure of the items is heavily skewed in all contextual bandit algorithms. These results signify that few items are frequently recommended at each round, while the majority of other items are either rarely or never recommended, indicative of significant exposure bias amplification in these algorithms. 

The second question to investigate is: is high disparity in the exposure of items due to the fact that those frequently recommended items are the high-quality ones and the rest of the under-exposed items are not enough qualified to be recommended, and as a result, the system simply recommended what users wanted? To answer this question, we need to look at the false positive rate (i.e., item was recommended, but was not clicked) of each item. The colorbar in Figure \ref{fig_npe_dist} shows the false positive rate for each item. This false positive rate for each item $i$ is computed as the number of times $i$ was recommended but was not clicked (i.e., false positive) divided by $E_i$ (i.e. false positive + true positive). 

As shown, there is no relationship between the false positive rate of the items and their exposure in all algorithms. That is, highly exposed items are not often selected or clicked by the users and also under-exposed items are not often ignored by the users. In \algname{CascadeLSB} on both datasets, there are cases where highly exposed items resulted in high false positive rate (red circle) and under-exposed items resulted in low false positive rate (blue circle). In \algname{CascadeLinUCB} and \algname{CascadeHybrid} on both datasets, although items made almost the same false positive rate, their exposure is different. These results show that skew in the representation of items in the recommendation lists cannot be due to the quality or relevancy of the items to the users' preferences, but can be an indicative of algorithmic bias.

In contrast to Figure \ref{fig_npe_dist} that item exposure is based on the appearance in the recommendation lists (i.e., equation \ref{npe}), Figure \ref{fig_npee_dist} shows the item exposure based on whether or not the item is examined (i.e., equation \ref{npee}). Again, the same pattern can be observed here: certain items are frequently shown on top of the list and have higher chance to be examined by the users even though they are not the most relevant ones (i.e., the ones with low false positive rate) in majority of cases. For instance, in \algname{CascadeLSB} on Amazon Book and \algname{CascadeLinUCB} on Movielens, highly exposed items resulted in high false positive rate. This means that those highly exposed items were frequently recommended to the users even though the users have not clicked on them. 

These results suggest that the studied contextual bandit algorithms tend to over-recommend certain items even though those items are not always selected or clicked by the users (high false positive rate), or under-recommend certain other items even though those items resulted in low false positive rate. These observations indicate that the degree of exposure given to the items is not based on the relevancy of the items to the users' interests and hence signify exposure bias in these algorithms. That is, these algorithms tend to favor certain items by over-exposing them even though they are irrelevant, while ignoring certain other items by not sufficiently exposing them to the users even though they are relevant. In this paper, we aim at addressing this exposure bias issue in these algorithms and reducing the disparity in exposure given to the items, while maintaining the accuracy of the recommendations.


\section{Exposure-Aware Reward model}

Contextual bandit algorithms described in section \ref{cb} iteratively learn users' preferences from their feedback on recommended items: clicked items are rewarded and unclicked items are penalized. These rewards and penalizations are reflected by updating parameter $M$ for all examined items (i.e., both clicked and unclicked) in equations \ref{update_M} and parameter $B$  only for clicked items in equations \ref{update_B}. Parameter $M$ is the co-variance matrix of the item features and parameter $B$ is the vector of importance weight assigned to the item features for the target user indicating how much important each features of the item is for the target user. When a user clicks on an item, it means that the user is interested in the contents of that item which are represented as the item features. Hence, $B$ cumulatively computes the importance weight for item features based on user feedback on different items. 

The problem with the above formulation is that it does not take into account the position of the items when rewarding or penalizing the recommended items. This means that clicked items on top of the list would be equally rewarded as the clicked items at the bottom of the list. Clicks on highly exposed items (i.e., items on top of the list) can be due to either the item was easily accessible, or the item was interesting for the user. However, less exposed items (i.e., items at the bottom of the list) require effort from the user to examine many other items and most likely those less exposed items are of the highest importance and interest for the user. Therefore, clicked items at the bottom of the list need to be rewarded more than the ones on top of the list.

Analogously, unclicked items need to be penalized differently. The unclicked items on top of the list should be penalized more than unclicked items at the bottom of the list. When a highly exposed item is not clicked by the user, it signifies that the recommendation model wrongly considered that item as of highest interest for the user. Penalizing this item more than less exposed unclicked items (i.e., items at the bottom of the list) would correct the recommendation model to not show it again on top of the list in the future. 

To properly adapt the model to the user feedback, we propose an \textit{exposure-aware reward model} that updates parameter $B$ based on the position of clicked and unclicked items in the recommendation list. According to the cascade model, user $u$ examines the ordered recommendation list of size $K$ at round $t$ one-by-one from top to the bottom and clicks on the most attractive item, then stops examining the rest of the items. This way, the clicked item would be rewarded, items above the clicked item would be penalized, and the rest of the items would be considered unobserved (i.e., neither rewarded nor penalized). If $u$ does not click on any recommended item, then all recommended items would be penalized. Hence, for each examined item $i$ at position $k$, $B_t$ at round $t$ would be updated as follows:
\begin{equation}\label{update_b}
    B_t = B_t + \mathcal{F}(C_{t,u},k,\gamma) . x_i
\end{equation}
\noindent where function $\mathcal{F}(C_{t,u},k,\gamma)$ returns the importance weight assigned to the features of item $i$ and can be defined as:
\begin{equation}\label{reward_model}
    \mathcal{F}(C_{t,u},k,\gamma) =
      \begin{cases}
        log_2(1+k) & \text{$k=C_{t,u}$} \\
        -\gamma \times \frac{1}{\log_2(1+k)} & \text{$k \textless C_{t,u}$}
      \end{cases}
\end{equation}
\noindent where $C_{t,u}$ is the index of the clicked item by target user $u$. As described in section \ref{exposure}, the term $1/\log_2(1+k)$ assigns an exposure value to each position in the list: higher value to the top position and lower value to the bottom position. Thus, in equation \ref{reward_model}, when $k=C_{t,u}$, then item at position $k$ would be rewarded: lower $k$ value (i.e., clicked item on top of the list) would result in lower reward and vice versa. On the other hand, when $k \textless C_{t,u}$, the item at position $k$ would be penalized: lower $k$ value would result in higher penalization and vice versa. $\gamma$ is a hyperparameter that controls the degree of penalization for unclicked items. Since there are much more unclicked items than the clicked ones, small $\gamma$ value allows to penalize unclicked items slightly and the algorithm mainly focuses on the clicked items to learn the users' preferences based on the preferred items. We set $\gamma=5e-5$ for the experiments.



\section{Experimental Evaluation of Exposure-Aware Reward Model}\label{experiment}

In this section, we perform sets of experiments and investigate the effectiveness of the proposed exposure-aware reward model in mitigating exposure bias.

\subsection{Experimental Setup}

To evaluate the effectiveness of the proposed exposure-aware reward model when incorporating it into the existing contextual bandit algorithms, we use the following metrics:
\begin{itemize}
    \item \textit{Clicks.} This metric measures the number of clicks observed on the recommendation lists generated by linear cascading bandit algorithm. We consider this metric as the accuracy of the recommendation model. Higher value for this metric indicates that the recommendation model is more accurate.
    \item \textit{Equality of Opportunity ($EO$).} This metric measures the equality of chance for all items to appear in the recommendation lists. Given the exposure distribution of items in recommendation lists of $T$ rounds, the goal of this metric is to see how uniform this distribution is. To measure the uniformity of the distribution, Gini Index \cite{vargas2014improving} is used and computed as:
    \begin{equation}
        Gini = \frac{\sum_{i=1}^{|\mathcal{I}|}{(2k-|\mathcal{I}|-1)NPE_i}}{|\mathcal{I}|-1}
    \end{equation}
    \noindent where items are indexed from 1 to $|\mathcal{I}|$ in $NPE_i$ non-descending order. Uniform distribution will have Gini index equal to zero which is the ideal case (lower Gini Index is better).
    \item \textit{Equality of Impact ($EI$).} This metric measures the equality of chance for all items to be examined by the users (higher chance to be clicked). That is, how uniform the distribution of examined items is. Similar to $EO$, Gini Index can be used to compute the uniformity of the distribution of the examined items, but instead of $NPE_i$, $NPEE_{i}$ is used for this computation. 
    \item \textit{Item coverage (IC).} This metric measures the fraction of items that appear at least once in the recommendation lists of $T$ rounds.  
\end{itemize} 

Contextual bandit algorithms used for the experiments in this paper involve hyperparameter $c$ that controls the degree of exploration. 
We perform a grid-search over $c=\{0.01,0.25,0.5,1\}$ to find the best-performing results. We used McNemar's test to evaluate the significance of results. For the rest of the paper, we show \algname{CascadeLSB}, \algname{CascadeLinUCB}, and \algname{CascadeHybrid} algorithms with the proposed exposure-aware reward model as \algname{EACascadeLSB}, \algname{EACascadeLinUCB}, and \algname{EACascadeHybrid}, respectively.

\input{tbl1}

\subsection{Comparison of best performing results}

Table \ref{res} shows the experimental results for the best performing hyperparameter setting for all algorithms in terms of number of clicks. On both datasets, the selected hyparameters are $c=0.01$ for \algname{LSB} and \algname{Hybrid} models, $c=1$ for \algname{LinUCB} models. These results compare the original cascading bandits with the exposure-aware cascading bandits when the algorithms are optimized to achieve the highest possible accuracy (clicks).

The results show that, in most cases, the exposure-aware contextual bandits improve both the number of clicks and exposure fairness for items compared to the original algorithms. In particular, the improvements achieved by \algname{EACascadeLinUCB} on both datasets is significant.

\begin{figure*}[t!]
    \centering
    \begin{subfigure}[b]{0.7\textwidth}
        \includegraphics[width=\textwidth]{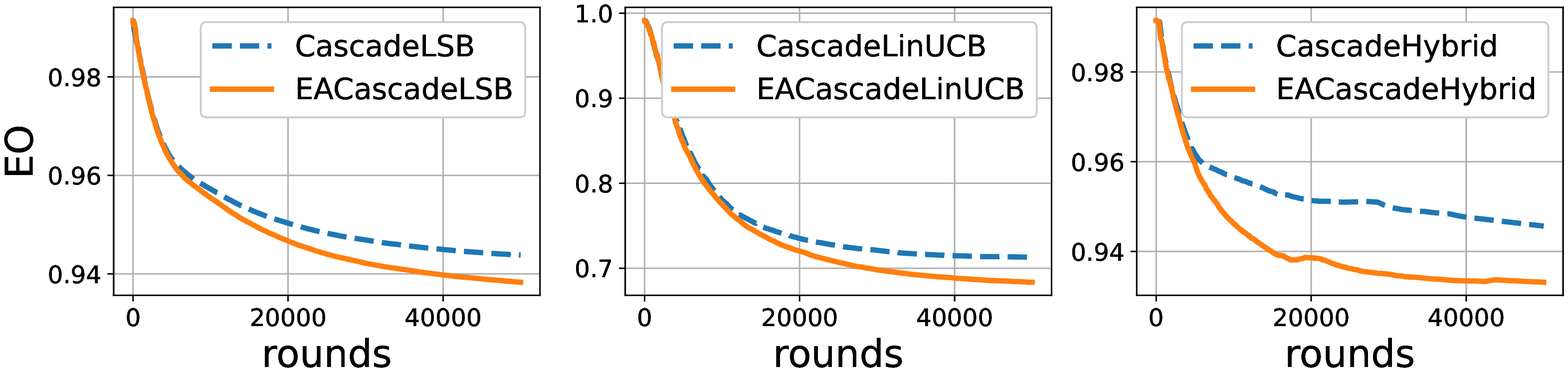}
        \caption{Amazon Book}\label{}
    \end{subfigure}
    \begin{subfigure}[b]{0.7\textwidth}
        \includegraphics[width=\textwidth]{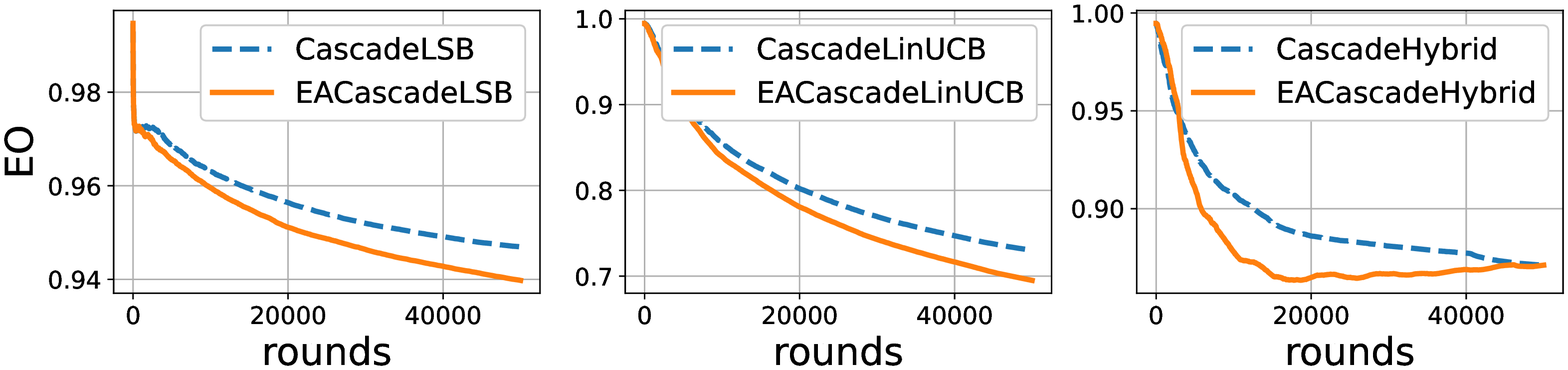}
        \caption{MovieLens}\label{}
    \end{subfigure}
\caption{Equality of opportunity ($EO$) at each round on Amazon Book and MovieLens datasets (lower $EO$ indicates higher exposure fairness).}\label{eo_per_round}
\end{figure*}

\begin{figure*}
    \centering
    \begin{subfigure}[b]{0.7\textwidth}
        \includegraphics[width=\textwidth]{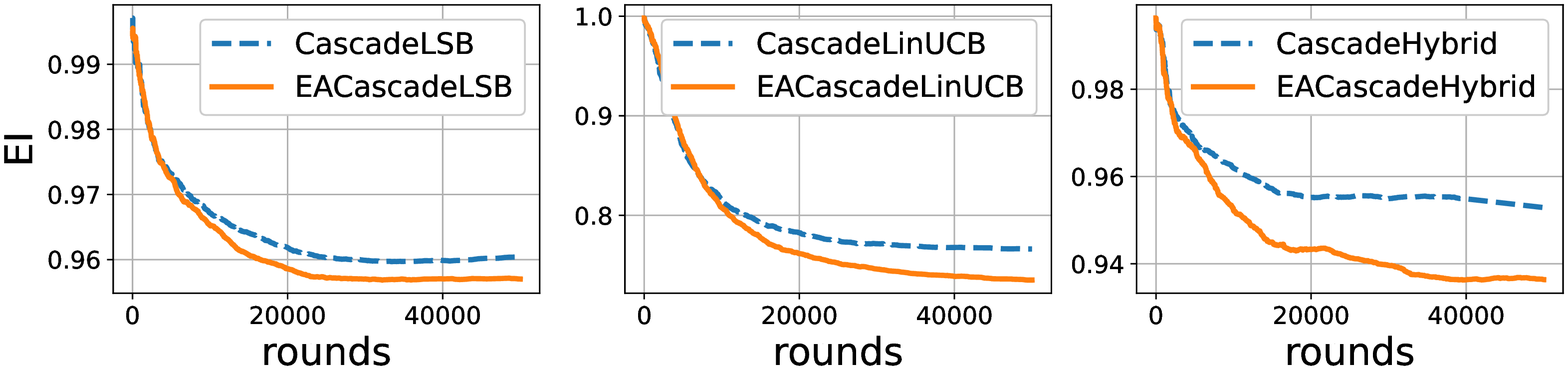}
        \caption{Amazon Book}\label{}
    \end{subfigure}
    \begin{subfigure}[b]{0.7\textwidth}
        \includegraphics[width=\textwidth]{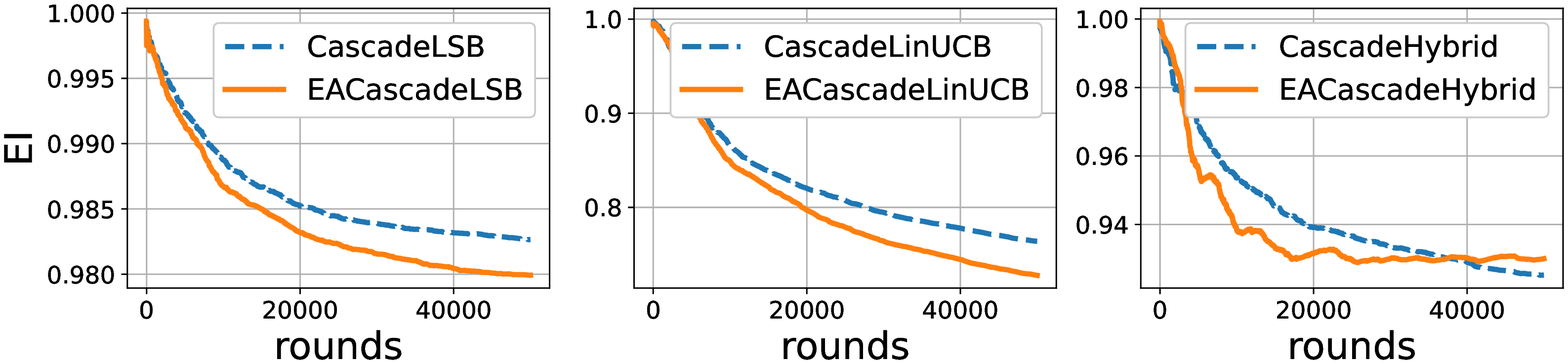}
        \caption{MovieLens}\label{}
    \end{subfigure}
\caption{Equality of impact ($EI$) at each round on Amazon Book and MovieLens datasets (lower $EI$ indicates higher exposure fairness). }\label{ei_per_round}
\end{figure*}

\subsection{Bias amplification mitigation}

Figure \ref{eo_per_round} shows $EO$ for each round yielded by each contextual bandit algorithm on both datasets. In these plots, at each round $t$, Gini Index is computed over the exposure of each item in recommendation lists ($NPE$) from round 1 to round $t$. The same computation is also used for computing $EI$ in Figure \ref{ei_per_round}, but instead of $NPE$, $NPEE$ is used for computing the exposure of items. For all cases, the results with the lowest $EO$ and $EI$ are reported, $c=0.01$ for all algorithms.  

Results in Figures \ref{eo_per_round} and \ref{ei_per_round} show how fairly exposure is distributed among items over time as the system is operating. Lower $EO$ and $EI$ at each round indicates that fairer exposure is achieved until that round. As shown, exposure-aware contextual bandits result in lower $EO$ and $EI$, fairer exposure, than original algorithms on both datasets in the long run. Also, from the plots, it is evident that if the system operates for more rounds, the disparity in $EO$ and $EI$ between the original cascading bandits and the proposed exposure-aware cascading bandits in all cases, except for \algname{EACascadeHybrid} on MovieLens, would be even higher, indicative of higher fairness achieved by the proposed exposure-aware cascading bandits compared to the original cascading bandits.

\section{Related work}

The problem of \textit{bias} and \textit{unfairness} in recommender systems is well-studied in the literature \cite{geyik2019fairness,mehrotra2018towards,zehlike2017fa}. It has been shown that recommendation algorithms suffer from various types of biases \cite{chen2020bias,baeza2020bias,chen2021bias}. One of these biases is \textit{exposure bias} which we studied in this paper.

Singh and Joachims in \cite{singh2018fairness} proposed a linear programming approach with fairness constraints that optimizes to achieve the maximum exposure fairness for items, with minimum loss in the relevance of the recommendations for the users based on a threshold. The authors showed that their approach is general and various notions of exposure fairness can be defined by this approach. In another work in \cite{singh2019policy}, the authors addressed the issue of exposure bias in stochastic ranking problems and proposed a new learning-to-rank algorithm. Their algorithm learns a policy over a distribution of rankings with fairness constraints to accurately generate recommendations while achieving the fairness objectives. 

In some other works, exposure bias is addressed by mitigating \textit{popularity bias} in input data: few popular items received the majority of interactions from the user, while majority of the items are rarely interacted \cite{ciampaglia2018algorithmic,abdollahpouri2020connection}. It has been shown that this skew in the representation of items in input data leads to over-recommendation of few popular items and under-recommendation of the majority of non-popular items. Zhu et al. in \cite{zhu2021popularity} studied popularity bias in dynamic recommendation setting and showed how recently introduced popularity-opportunity \cite{zhu2021popularityopportunity} bias can lead to bias amplification in long-run. The authors proposed a false positive correction method for matrix factorization model which mitigates bias by reducing the number of false positive in recommendation results.

In addressing exposure bias in dynamic recommendation setting, in particular in contextual bandit algorithms, Mansoury et al. \cite{mansoury2021unbiased} modified the recommendation generation step by incorporating a discounting factor into the score computation for items. This discounting factor dynamically adjusts the score for the items according to their exposure in the past, meaning that if two items have almost the same relevance score, then it allows the item with low exposure to appear in the recommendation list. Wang et al. in \cite{wang2021fairness} proposed the notion of fairness regret and reward regret to learn an optimal policy that ensures merit-based exposure. In a more recent work, Jeunen and Goethals in \cite{jeunen2021top} studied the merit-based fairness in contextual bandit algorithms and proposed a tolerance parameter for improving fairness. This parameter allows for randomizing the order of certain items in top-$K$ recommendation list that results in maximal increase in fairness with minimum loss in reward. In fact, it attempts to find the position $K'$ in top-$K$ recommendation list that randomizing the order of items from the first position to the $K'$-$th$ position would lead to loss in expected clicks at most up to the threshold $\epsilon$. This would rerank the recommendation lists generated by the original bandits, modifying the recommendation generation process at line 4 of Algorithm \ref{alg:cap}. However, our idea in this paper is modifying the reward model in line 6 on Algorithm \ref{alg:cap}. One drawback of \cite{jeunen2021top} is that finding the position $K'$ requires computing the expected clicks for different permutations of $K$ recommended items which makes this approach computationally expensive. However, our proposed exposure-aware reward model does not require any additional computations. 

In contrast to these works which adopt the notion of merit-based fairness, we study equality of fairness for items in recommendation results. Also, in contrast to \cite{wang2021fairness} which studies the general single-armed contextual bandits, our work in this paper focuses on fairness of exposure in top-$K$ contextual bandits algorithms.  

\section{Discussion and Future work}

An interesting future direction is conducting research on exposure fairness from user perspective. In this paper, we studied how recommendation models fairly distribute the total exposure among all the items, but it is unclear how this fair distribution of exposure would affect different users. Considering item coverage as a metric for measuring exposure fairness, for example, few users may be exposed with many unique items in their recommendation lists, while majority of other users may repeatedly be exposed with few unique items. In this situation, although the recommendation model covered majority of the items in the recommendation lists (high item coverage), the average item coverage for each user is still very low. Analogously, other notions of exposure fairness introduced in this paper ($EO$ and $EI$) can be considered for user-centered view. 

In another future work, we plan to extend our analysis to other classes of bandit algorithms, such as those based on Thompson Sampling and compare the effectiveness of the proposed exposure-aware reward model in those algorithms with linear cascading bandits studied in this paper.


\section{Conclusion}

In this paper, we studied the problem of exposure bias in a specific class of contextual bandit algorithms called \textit{Linear Cascading Bandits}. We observed that these algorithms fail to provide a fair chance for certain items to appear in the recommendation lists. That is, certain items are over-recommended even though those items are not selected by the users, resulting in under-recommendation of majority of other items. To address this problem, we proposed an \textit{exposure-aware} reward model that rewards clicked items and penalizes unclicked items based on both user feedback and the position of the items in the recommendation list. The proposed model allows for properly adapting to the user feedback by downgrading the unclicked items and promoting clicked items in the recommendation lists of subsequent rounds. Experiments using thee contextual bandit algorithms on two real-world datasets showed that the proposed exposure-aware reward model is effective in mitigating exposure bias while maintaining the recommendation accuracy. 

\section*{Acknowledgement}
This project was funded by Elsevier’s Discovery Lab.

\bibliographystyle{ACM-Reference-Format}
\bibliography{ref}
\end{document}

%% file: tbl1.tex
\captionsetup[table]{skip=4pt}
\begin{table}[t]
\aboverulesep=0ex 
\belowrulesep=0ex 
\footnotesize
\centering
\setlength{\tabcolsep}{3pt}
\captionof{table}{Comparison of best performing results in terms of number of clicks for original and exposure-aware contextual bandits after $T$ rounds of running recommendations. The bolded entries show the best values and the underlined entries show the statistically significant change with
$p<0.1$.} \label{res}
\begin{tabular}{l|rrrr|rrrr}
\midrule
\midrule
 \multirow{2}{*}{algorithms} & \multicolumn{4}{c|}{Amazon Book} & \multicolumn{4}{c}{MovieLens} \\\cline{2-5}\cline{6-9}
 
 & $clicks$ & $EO$ & $EI$ & $IC$ & $clicks$ & $EO$ & $EI$ & $IC$ \\
 \midrule

 \algname{CascadeLSB} & 11,255 & 0.955 & 0.968 & 0.062 & \textbf{33,642} & 0.988 & 0.995 & 0.567 \\
 \algname{EACascadeLSB}  & \textbf{11,837} & 0.955 & \textbf{0.966} & \textbf{0.069} & 32,341 & \textbf{0.983} & \textbf{0.992} & \textbf{0.592} \\
 \hline
 
 \algname{CascadeLinUCB}  & 16,283 & 0.713 & 0.766 & 0.921 & 15,706 & 0.731 & 0.765 & 0.944 \\
 \algname{EACascadeLinUCB}  & \textbf{16,455} & \underline{\textbf{0.683}} & \underline{\textbf{0.734}} & \textbf{0.933} & \textbf{15,970} & \underline{\textbf{0.695}} & \underline{\textbf{0.727}} & \textbf{0.953} \\
 \hline
 \algname{CascadeHybrid}  & 10,238 & 0.955 & 0.962 & 0.792 & \textbf{31,491} & \textbf{0.972} & 0.994 & 0.350 \\
 \algname{EACascadeHybrid}  & \textbf{10,739} & \underline{\textbf{0.885}} & \underline{\textbf{0.890}} & \underline{\textbf{0.813}} & 30,345 & 0.973 & \textbf{0.991} & \underline{\textbf{0.424}} \\
 \midrule
\midrule
\end{tabular}
\end{table}